\documentclass[11pt,preprint]{aastex}
\usepackage{epsf}
\usepackage{psfig}
\usepackage{graphicx}
\begin{document}
\title{Aplicando modelos de granos de polvo con propiedades
meteor\'iticas al continuo ionizante de algunos N\'ucleos  Activos de
Galaxias.}

{ \it``Coloquio: Origen y estructura del Sistema Solar'', Revista
  de Ciencias Geológicas del Instituto de Geolog\'ia, 2007}

\author{ Sinhu\'e Amos Refugio Haro-Corzo$^1$ \& Luc Binette$^2$
\affil{$^1$ Instituto de Ciencias Nucleares, Universidad Nacional Aut\'onoma
de M\'exico, Apartado Postal 70-543, C.P. 04510, Ciudad de M\'exico.}
{$^2$ Instituto de Astronom\'ia, Universidad Nacional Aut\'onoma
de M\'exico, Apartado Postal 70-264, C.P. 04510, Ciudad de M\'exico.}}

\begin{abstract}
En este trabajo, presentamos las diversas curvas de extinci\'on que
hemos calculado con base a la teor\'ia Mie para granos de polvo
esf\'ericos, con distribuciones de acuerdo a su tama\~no (entre
50-2500\,\AA) y con diversos compuestos qu\'imicos hallados en
meteoritos. Bajo el posible paradigma que hemos desarrollado, estas
curvas de extinci\'on al ser aplicadas a un continuo ionizante
te\'orico y comparadas con 11 espectros observados de quasares, hemos
encontrado de manera exitosa reproducir tanto el {\it quiebre UV}
as\'i como el {\it problema de suavidad} presentes en el continuo
ionizante de algunos N\'ucleos Activos de Galaxias.

We construct extinction curves based on Mie's theory for spherical
dust grain with size distributions between 50-2500\,\AA and composed
by elements found in meteorites.  We explore the issue of whether the
observed spectral energy distribution might be dust absorbed in the
far and near-UV by this kind of dust grains. Using this
approach, satisfactory fits to the 11 spectra can be obtained.

\end{abstract}

\section{Introducci\'on}
Observaciones espectrosc\'opicas realizadas con diversos telescopios
han encontrado que la Distribuci\'on Espectral de Energ\'ia Ionizante
(SED) de los quasares (subgrupo de los N\'ucleos Activos de Galaxias)
presentan los siguientes rasgos en el dominio ultravioleta (UV):
Emisi\'on de radiaci\'on (continuo ionizante) proveniente de la
regi\'on central y con superposici\'on de l\'ineas de emisi\'on y
absorci\'on. A su vez, el continuo ionizante presenta un {\it quiebre
  UV}, es decir, un cambio abrupto de pendiente alrededor
1100\,\AA\ \citep[][de aqu\'i en adelante H07]{haro07} y un {\it
  problema de suavidad}, es decir, dispersi\'on entre los
\'indices espectrales observados en el UV lejano \citep{binette07h}. A
la fecha es un tema abierto que exige soluci\'on para entender la
naturaleza de los hoyos negros supermasivos que viven en los quasares.


A lo largo de nuestra investigaci\'on, hemos desarrollado el siguiente
paradigma que resuelve exitosamente el {\it quiebre UV} as\'i como el
{\it problema de suavidad}: existe una SED te\'orica universal que emerge del
quasar, la cual es absorbida por cantidades apropiadas de (1)
gas/polvo intr\'inseco y (2) el medio intergal\'actico y Gal\'actico.
\citet{binette07h} propusieron diversos escenarios para explicar el punto
(1), por lo que en este trabajo s\'olo profundizaremos en este mismo
punto, es decir, exploraremos la absorci\'on debida a dos pantallas de
polvo que est\'an cerca y alrededor al quasar, donde cada pantalla de
polvo est\'a formada por un s\'olo compuesto qu\'imico (silicatos o
carb\'on amorfo o nanodiamantes). As\'i, cada pantalla tiene una curva
de extinci\'on peculiar que relaciona la secci\'on eficaz del material
absorbente a cada longitud de onda.

\section{Curvas de Extinci\'on del Polvo}
En esta secci\'on explicamos los c\'alculos realizados para construir
los factores de eficiencia de la extinci\'on (Q$_{ext}$), los cuales
son necesarios para formar la secci\'on eficaz. Estos c\'alculos de
Q$_{ext}$ est\'an basados en la teor\'ia Mie \citep{bohren83} para una
distribuci\'on ($n(a) \propto a^{+\beta}$) de granos esf\'ericos de
radio ``a'' entre a$_{min}$=50 y a$_{max}$=2500\,\AA, \'indice
$\beta=-3.5$ \citep{bookwhittet}, isotr\'opicos, homog\'eneos y con
varias composiciones qu\'imicas, i.e. con varios \'indices complejos
de refracci\'on (m=n$-i$k) acorde al material \citep[m\'as detalles en][]{harocorzotesis}.

Para construir  las secci\'on eficaz a partir de Q$_{ext}$ se
resuelve la siguiente ecuaci\'on para la distribuci\'on de granos:
$\sigma_{dust} =  C_{gr}  {\int_{a_{min}}^{a^{max}} d{\rm ln}a} \; \pi a^{3+\beta} Q_{ext}(a,\lambda,n,k)$ 
, donde la constante de normalizaci\'on $ C_{gr}$ y el volumen 
del grano esf\'erico $V_{gr}$ son calculados con: 
$C_{gr} = \frac{Z_{el} \; \mu_{gr} \; m_H}{ \rho_{gr} \, V_{gr}}
\,\,\, ;\,\,\, V_{gr} = {\int_{a_{min}}^{a_{max}} d{\rm ln}a} \; \frac{4}{3}\pi a^{4+\beta}$
, donde $Z_{el}$ es la abundancia num\'erica del elemento a considerar
con respecto al hidr\'ogeno (H), $m_H$ es la masa del \'atomo del H,
$\rho_{gr}$ es la densidad de masa del grano (g\,cm\,$^{-3}$),
$\mu_{gr}$ el peso molecular promedio del grano. Adem\'as, debido a
que pocos quasares se les ha determinado su abundancia qu\'imica, hemos
asumido como primera aproximaci\'on abundancias solares.

Una vez esto, la t\'ecnica usada para reproducir el continuo ionizante
observado (F$_{obs}$) y explorar los efectos de las diversas pantallas
de polvo es: partir de una SED te\'orica universal (F$_{int}$) en
forma de joroba y la cual es multiplicada por la transmisi\'on debida
al polvo (absorci\'on del polvo en funci\'on de su secci\'on eficaz
$\sigma_{dust}$ [cm$^{2}$] y de la densidad columnar equivalente
N$_H$[cm$^{-2}$], como un par\'ametro libre),
i.e. F$_{obs}$=F$_{int}\times exp(-N_H\,\sigma_{dust})$. Con este
planteamiento buscamos mejorar el ajuste al continuo en el dominio del
UV y adem\'as buscar la relaci\'on con el espectro en los rayos X para
cada quasar. Con base en esto, ahora s\'olo falta definir la
composici\'on qu\'imica del polvo.

\subsection{Pantalla de polvo compuesta por Nanodiamantes}

Con el objetivo de corregir el {\it quiebre UV} hemos seguido el
trabajo de \citet[de aqu\'{\i} en adelante B05]{binette05}, quienes
exploraron diversas pantallas de polvo compuestos por diamantes de
tama\~nos nanom\'etricos ($10^{-9}$\,m). Los nanodiamantes est\'an
compuestos por \'atomos de carb\'on colocados en una red cristalina y
en algunas ocasiones presentan impurezas en la superficie. La
abundancia solar del carb\'on es de $3.6310^{-4}$ con respecto al
hidr\'ogeno y entonces el peso molecular es de 12 para todos los
granos.

\subsubsection{Nanodiamantes Meteor\'itico: modelos A}

\begin{figure}[!ht]
\begin{center}
\includegraphics[width=8.5cm,keepaspectratio=true]{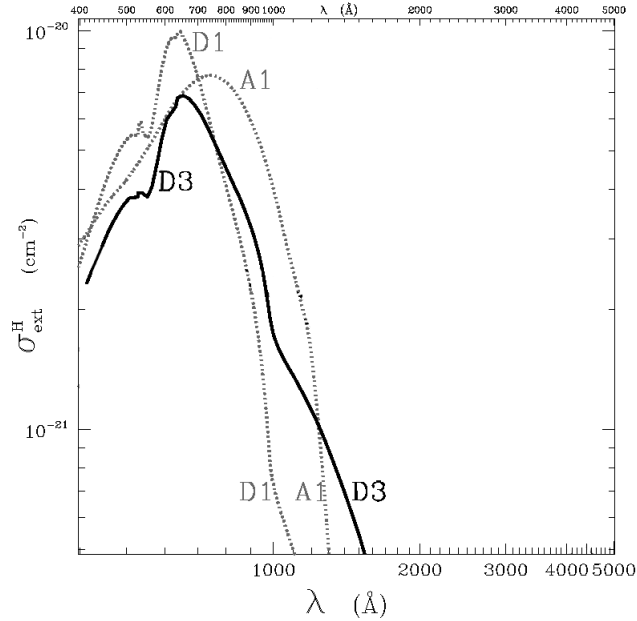}
\caption{Curvas de extinci\'on o secci\'on eficaz de absorci\'on como
funci\'on de la la longitud de onda para granos de polvo compuestos
por nanodiamantes. Aqu\'{\i} ilustramos las tres curvas que hemos utilizado
a lo largo de nuestra investigaci\'on para reproducir el {\it quiebre
UV} a partir de la absorci\'on por granos de polvo. Las tres curvas
var\'{\i}an de acuerdo a su composici\'on como sigue: nanodiamantes
meteor\'itico (A1), nanodiamante c\'ubico (D1 y D3). En especial, hemos
adoptado a la curva D3 (l\'inea continua), que est\'a en el r\'egimen
de granos pequen\~os entre 3 y 200\,\AA, como la representante m\'as
adecuada para resolver el problema del {\it quiebre UV}.  }
\end{center}
\label{fig:ND}
\end{figure}

Uno de los ejemplos m\'as estudiado es el meteorito de Allende, el
cual aterriz\'o en la ciudad de Allende en el estado de Chihuahua,
M\'exico, el 8 Febrero de 1969. En este meteorito fueron encontrados
nanodiamantes con impurezas de hidr\'ogeno\footnote{Hacemos notar que
tambi\'en existen nanodiamantes con impureza de N en su superficie,
los cuales ser\'an explorados en trabajos futuros} a lo largo de la
superficie. Estos nanodiamantes tiene tama\~nos t\'{\i}picos entre 10
y 100\,\AA, densidad de 2.3\,g\,cm\,$^{(-3)}$ y bandas en emisi\'on
en el IR, i.e, a 3.43 y 3.53\,$\mu$, las cuales fueron identificadas en
algunas estrellas \citep{VT02}.\\ El nanodiamante es el material
granular m\'as abundante en meteoritos primitivos \citep{lb87, lm87,
z98}.  La sobreabundancia de los nanodiamantes con respecto a las
dem\'as componentes del meteorito, sugiere que los granos del medio
interestelar fueron acretados directamente en planetisimales y los
cuales permanecen relativamente inalterados en los n\'ucleos de los
asteroides.\\ Dada la abundancia y dureza de los nanodiamantes,
\citet{mutschke04} pudieron aislarlos del resto del meteoro con
disolventes qu\'imicos para luego investigar los \'indices de
difracci\'on complejos, lo cual permiti\'o a B05 construir la secci\'on
eficaz {\bf A1} (ilustrada en la figura\,\ref{fig:ND}) y reproducir
por primera vez el {\it quiebre UV}.

\subsubsection{Nanodiamantes C\'ubico: modelos D}
Este tipo de nanodiamante no tiene impureza en la superficie. La
densidad de granos es de 3.51\,g\,cm\,$^{-3}$ \citep{ep85}. La
secci\'on eficaz de los nanodiamantes c\'ubicos fue extrapolada hacia
el UV lejano, dado que no encontramos valores de laboratorio de los
\'indices de refracci\'on por debajo de 413\,\AA.\\ La curva de
extinci\'on utilizada por B05 fue etiquetada como {\bf D1}, la cual
est\'a en el {\it r\'egimen de granos peque\~nos}, i.e, en el
intervalo de tama\~nos entre a$_{min}= 3$ y a$_{max}=25$\,\AA, en
cambio la curva de extinci\'on {\bf D3} tiene un intervalo de
tama\~nos m\'as amplio, es decir, entre a$_{min}= 3$ y
a$_{max}=200$\,\AA.

En la figura\,\ref{fig:ND} comparamos las curvas de extinci\'on
calculadas para los diferentes tipos de nanodiamantes. Hacemos notar
que el pico de la secci\'on eficaz para D1, A1 y D3 ocurre a la
longitud 640, 741 y 640\,\AA\ respectivamente. Esta particularidad de
la secci\'on recta de los nanodiamantes, junto con el abrupta
ca\'{\i}da de la secci\'on eficaz hacia el rojo de 1000\,\AA\ son
rasgos \'unicos de este tipo de polvo, los cuales dejan una clara
huella en el espectro UV bajo el paradigma del trabajo presente (ver
H07). As\'i, los modelos con polvo que usan componentes altamente
ordenadas (por ejemplo los modelos D) tienden a producir quiebres
abruptos y bien localizados, mientras que el polvo dopado con
componentes desordenadas (por ejemplo A) tienden a producir quiebres
anchos.

\subsection{Curva de extinci\'on tipo SMC}

\begin{figure}[!ht]
\begin{center}
\includegraphics[width=8.5cm,keepaspectratio=true]{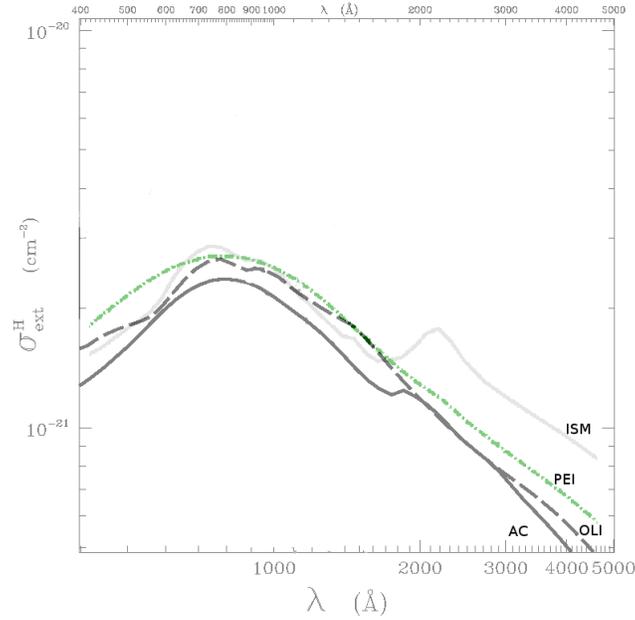}
\caption{Curvas de extinci\'on como funci\'on de la longitud de onda
  para diferentes compuestos de granos de polvo. Por comparaci\'on
  hemos superpuesto la curva de extinci\'on tipo medio interestelar
  Gal\'actica (l\'inea continua-tenue y etiquetada como ISM). El rasgo
  caracter\'{\i}stico de esta curva es el m\'aximo (joroba en
  absorci\'on ancha y sim\'etrica) que esta a 2175\,\AA\ y que es
  atribuida a la presencia de grafito. Las curvas de extinci\'on tipo
  SMC utilizadas para resolver el problema de suavidad en los quasares
  est\'an ilustradas como sigue: La curva de extinci\'on calculada por
  \citet{pei92} (l\'inea punto-segmento y etiquetada como ``PEI'')
  est\'a escalada por un factor 5.5 para facilitar la comparaci\'on
  con el resto de las curvas. La curva que representa a los granos de
  polvo con propiedades del olivino ($MgFeSiO_4$) est\'a ilustrada
  como una l\'inea segmentada, etiquetada como ``OLI'' y escalada por
  un factor de 2. La curva que ilustra la absorci\'on que presentan
  los granos de polvo compuestos por Carb\'on Amorfo est\'a ilustrada
  como una l\'inea s\'olida y etiquetada como ``AC''. Est\'a \'ultima
  curva no necesita ser escalada y adem\'as presenta un rasgo peculiar
  alrededor de 1700\,\AA, as\'i esta es la curva adoptada para
  representar la extinci\'on tipo SMC.}
\end{center}
\label{fig:SMC}
\end{figure}

\citet{richards03} infirieron curvas de extinci\'on
similares al tipo Nube Menor de Magallanes (SMC) a partir de los miles
de espectros de quasares provenientes del catastro con el Sloan
(SDSS). A la fecha, en los congresos internacionales es tema de debate
la forma, la inclinaci\'on, la metalicidad y los rasgos caracter\'{\i}sticos
de la curva de extinci\'on inferida en los objetos extragal\'acticos
con corrimiento al rojo moderado. Esto nos se\~nala que los objetos
est\'an envueltos en medios ambientes distintos y que requieren una
investigaci\'on m\'as detallada. En el proyecto que hemos estando
desarrollando, asumimos que la curva de extinci\'on aplicable a los
quasares es tipo SMC, ya que en general las diversas curvas en el
dominio UV son similares, como es mostrado en la
figura\,\ref{fig:SMC} donde comparamos la curva de extinci\'on tipo
SMC (l\'inea segmento-punto) con la Gal\'actica (l\'inea continua
tenue y etiquetada como ISM). Es notable en la comparaci\'on que la
curva tipo SMC tiene una pendiente muy pronunciada en el cercano UV
($<1100$\,\AA) y no tiene la joroba en absorci\'on a 2175\,\AA, que
caracteriza a la curva de extinci\'on ISM. Basados en lo anterior y
con el objetivo de resolver el {\it problema de suavidad} en el
intervalo del cercano UV, hemos tratado de reproducir la curva de
extinci\'on tipo SMC con los siguientes tres compuestos.

\subsubsection{Silicatos}
La primera curva que construimos (l\'inea segmento-punto y etiquetada
como PEI) se baso en la curva de \citet{pei92}, quien emple\'o granos
de polvo compuestos por silicatos para reproducir la curva de
extinci\'on deducida por
\citet{prevot84} a partir de los datos de la SMC. Sin embargo, este
modelo requiere de 2 a 3 veces m\'as cantidad de silicio disponible en
el medio interestelar de la SMC. Adem\'as es necesario escalar la
curva por un factor 5.5 para hacerla comparable con la curva tipo
ISM. As\'i, debido a que las abundancias utilizadas para generar esta
curva de extinci\'on no son f\'isicamente compatibles con las
observadas, descartamos este tipo de granos de polvo.
\\ En la segunda curva asumimos que el polvo est\'a formado por silicato
tipo olivino (MgFeSiO$_4$). Los \'atomos de Mg y el Fe son
aproximadamente igualmente abundantes en el medio interestelar. Ambos
residen principalmente en el polvo interestelar
\citep{draine03xr} y en los meteoros
\citep{delaney00}. Es entonces razonable considerar granos de polvo
compuestos por MgFeSiO$_4$, aunque es posible, que por ejemplo, el Fe
puede tener otra composici\'on qu\'imica. La densidad del olivino es
500 g\,cm$^{-3}$, los \'indices complejos de refracci\'on fueron
extra\'{\i}dos de \citep{draine03xr} y las abundancias solares son (en
escalas de $10^{-5}$): 3.8 para el Mg, 4.68 para el Fe, 3.5 para el Si
y 86.1 para el O. As\'i el peso molecular del grano es de 172.19. La
curva de extinci\'on del olivino est\'a ilustrada en la
figura\,\ref{fig:SMC} (l\'inea segmentada-larga y etiquetada como
OLI), la cual presenta un factor 2 de normalizaci\'on por debajo de
los datos.

\subsubsection{Carb\'on Amorfo AC } 

Este es un arreglo no cristalino de los \'atomos de carb\'on, que
forman anillos unidos aleatoriamente, por lo que entre estos anillos
se forman huecos llamados poros. Los \'indices de refracci\'on
complejos as\'i como la densidad del carb\'on ($\rho
=1.85\,$g\,cm\,$^{-3}$) fueron extra\'{\i}dos de la tabla\,1
(columna\,2 y 4) de \citet{rm91}.  El peso molecular que adoptamos es
$\mu_{gr} = 12$, la abundancia solar es Z$_c=3.63 \times 10^{-4}$ con
respecto al hidr\'ogeno y asumimos que todo el carb\'on est\'a oculto
dentro de los granos de polvo. El intervalo \'optimo del tama\~no de
los granos es entre a$_{min}\,=\,50$\,\AA\ y
a$_{max}\,=\,1400$\,\AA\ con una resoluci\'on de 0.01\,\AA. Afinamos
el tama\~no de los granos hasta que reprodujera la $forma$ de la curva
de extinci\'on tipo SMC. La curva de extinci\'on resultante est\'a
mostrada en la figura\,\ref{fig:SMC} en l\'inea continua y etiquetada
como AC. \\ Asumimos que la curva AC es la curva de extinci\'on m\'as
atractiva para representar la extinci\'on tipo SMC, debido al hecho
que no requiere renormalizaci\'on y adem\'as est\'a presente un rasgo
alrededor de 1700\,\AA\ que es observado en algunos espectros de
quasares.

\section{Discusi\'on}
En los quasares, el continuo ionizante observado presenta un cambio de
pendiente abrupto alrededor de 1100\,\AA\ y adem\'as una pendiente
suave en el cercano UV, espectros que no pod\'ian ser ajustados
simult\'aneamente con los modelos que reproducen tanto el continuo
as\'i como las l\'ineas de emisi\'on. En esta investigaci\'on
encontramos que de las cinco posibles pantallas exploradas bajo el
paradigma de una SED intr\'inseca universal proveniente de la
m\'aquina central del quasar, nosotros favorecemos las pantallas
intr\'{\i}nsecas en absorci\'on compuestas por granos de polvo de
nanodiamante c\'ubico y carb\'on amorfo para explicar
satisfactoriamente (ver ajustes en H07) el {\it quiebre UV} y el {\it
  problema de suavidad}. Tambi\'en encontramos que en promedio la masa
de polvo necesaria para formar est\'as pantallas es de 0.003 masas
solares (B05), suponiendo que la pantalla de polvo est\'a a 1\,pc del
quasar.  \\ La t\'ecnica de proceder para cada quasar fue primero corregir
el espectro por el {\it quiebre UV} con la curva D3 y posteriormente
corregir el {\it problema de suavidad} con la curva AC. Lo anterior es
debido a que la curva D3 absorbe principalmente en la regi\'on del
 UV lejano y muy poco en el cercano UV. Aunque la curva D3
(nanodiamantes c\'ubicos) es una simplificaci\'on del caso planeado
por B05, quien utiliz\'o nanodiamantes meteor\'iticos as\'i como
c\'ubicos, la justificaci\'on para utilizar nanodiamantes est\'a
inspirada en que los nanodiamantes meteor\'iticos adem\'as de
presentar un quiebre agudo en su curva de extinci\'on, tambi\'en
tienen bandas de emisi\'on en la banda IR. Es necesario continuar
con esta exploraci\'on y en caso dado detectar la respectiva emisi\'on
en la banda del IR para los quasares analizados por H07. Recientemente
\citet{dediego07} descarto la contribuci\'on del polvo compuesto por
nanodiamante meteor\'itico en el quasar 3C\,298 al comparar la
predicci\'on de la emisi\'on IR de nuestro paradigma con los datos del
telescopio espacial Spitzer. Otra l\'inea de investigaci\'on es
averiguar si los mecanismos que permitieron la formaci\'on de
nanodiamantes en el sistema solar pudieron darse en los quasares. Por
\'ultimo, consideramos que las alternativas se\~naladas por
\citet{binette07h}, donde algunas no consideran nanodiamantes, ser\'an
valiosas para ser exploradas en trabajos posteriores.

\acknowledgements Este trabajo est\'a financiado por la beca
Postdoctoral CONACyT de S.A.R.H.C. El apoyo para L.B. proviene del
proyecto J50296 de CONACyT. Agradecemos la asistencia de Alfredo
D\'iaz y de Abraham Roldan.


\bibliography{sed}

\end{document}